\documentclass{elsart}

\usepackage{natbib}

\usepackage{epsfig}

\usepackage{amssymb}

\begin{document}

\begin{frontmatter}



\title{Spectra and power of relativistic jets}


\author{Gabriele Ghisellini}

\address{Osservatorio Astronomico di Brera, Via Bianchi 46, I--23807 Merate Italy}

\begin{abstract}
The power of blazar jets rivals the power that gravity can extract from
accreting matter.
The mechanism launching and accelerating jets can be considered
as the most efficient engine operating in radio--loud sources.
It is still a matter of debate if the jet carries this power to the radio
lobes, hundreds of kpc away, in the form of Poynting flux or bulk kinetic
energy, or both, and if these two ingredients have relative weights
changing along the way.
Accordingly, there are two (or more) possible general scenarios for 
how the jet can dissipate part of its power into radiation.
It can be through e.g. reconnection of the magnetic field in the
purely electromagnetic scenario, or through internal shocks in the 
matter dominated picture.
Ways to discriminate these ideas are welcome.

\end{abstract}

\begin{keyword}
Galaxies: jets; Quasars: general; BL Lacertae objects.


\end{keyword}

\end{frontmatter}

\section{Introduction}

The bulk of the radiation we observe from blazars is produced in 
a well localized region of their relativistic jets.
This is, in my opinion, one of the most important results
of the $\gamma$--ray observations performed by EGRET 
onboard the Compton Gamma Ray Observatory, and of the more recent
observations performed by ground based Cherenkov telescopes.
When observational campaigns were organized at $\gamma$--ray
and other frequency bands, they proved that also the X--ray
flux was varying at the same time of the $\gamma$--ray flux, 
even if with a smaller amplitude.
By demonstrating that both fluxes are likely to be
produced in the same region, this puts a strong limit
on the local photon energy densities, which must be 
sufficiently small to avoid the absorption caused by the 
photon--photon pair production process. 
This implies that the observed flux is enhanced by beaming 
and therefore that the source is moving relativistically.
The same transparency constraint requires also that 
the $\gamma$--ray emitting region of the jet is sufficiently
far from the X--ray emitting corona sandwiching the accretion disk.  
On the other hand, the hour--day variability timescales
poses an upper limit on the size of the emitting region.
The jet region where most of the dissipation takes place
must then be at a few hundreds of Schwartzchild radii from the 
black hole (Ghisellini \& Madau 1996)

The other important result of the high energy data is that we now 
know the entire spectral energy distribution (SED) of blazars: 
since the $\gamma$--ray luminosity is often dominating the total 
radiation output, we now know the total radiated power. 
This is large.
The total power in radiation only, accounting for beaming, equals 
$L_{\rm obs}/\Gamma^2$, where $L_{\rm obs}$ is the bolometric 
luminosity calculated assuming isotropy and $\Gamma$ is the bulk 
Lorentz factor.
Jets must carry a power larger than that, implying for instance 
that the jet of S5 0836+71 (which has $L_{\rm obs}\sim 10^{49}$ 
erg s$^{-1}$, Tavecchio et al. 2000), should carry more than 
$10^{47}/(\Gamma/10)^2$ erg s$^{-1}$.
The question is in what form.
It can be purely electromagnetic, as advocated by Blandford (2002) and 
by Lyutikov \& Blandford (2002), or largely in the form of bulk 
motion of matter, with a relatively weak Poynting flux.
Also, the relevant ``matter" can be protons, or instead mostly
electron--positron pairs.
And of course there can be all the combinations of the above
ingredients, possibly changing weight at different distances
from the black hole: this underlines the importance to
find ways to measure the power of jets and their
matter content at different scales.
For instance, one can calculate the amount of the emitting particles
and the magnetic field required to account for the radiation we see,
and if also the size and the bulk Lorentz factor are known,
one can set a lower limit to the jet power (it is a lower limit
because there can be other cold, non emitting, particles).
This can be done at the jet scale where most of the radiation 
is produced (i.e. $10^{17}$ cm from the black hole, see e.g. 
Ghisellini 1999 and Fig. 1), where superluminal motion is detected 
(i.e. at the parsec--scale, see e.g. Celotti \& Fabian 1991);
at the large jet scales (i.e. at hundreds of kpc, as recently 
observed by Chandra, VLA, HST, see e.g. Ghisellini \& Celotti 2001; 
Tavecchio et al. 2000); and finally at the scale of the radio
lobes, thought as calorimeters (e.g. Rawlings \& Saunders 1991).
These studies point to the important conclusion that the engine
generating the jet can be even more efficient than accretion.
This would become dramatically clear if also gamma--ray bursts
(GRBs) are jetted sources.

\section{Jet power}

Fig. 1 (left panel) shows an updated version of the distribution of
the jet power found in blazars (Celotti \& Ghisellini 2003, in prep.).
It is found, as mentioned before, by assuming a synchrotron
inverse Compton model for the emission, which fixes the number
of particles and the value of the magnetic field of the emitting
region, together with its dimension and the value of the bulk 
Lorentz factor.
The distribution of the emitting particles is assumed to extend
down to $\gamma_{\rm min} =1$, and to derive the power in protons 
we have assumed one proton per emitting electrons (i.e. no pairs).
We can see that the powerful blazars (i.e. the flat radio spectrum quasars,
FSRQs for short) have the largest power in emitted radiation.
The power carried by the relativistic electrons, $L_{\rm e}$, 
does not reach these levels.
This is a consequence of the rapid cooling suffered by the emitting
particles in FSRQs, which is shorter than the light crossing time:
since they emit so rapidly, they have small average random energies,
and their relativistic inertia is relatively small.
Consider also that the relativistic electrons must be accelerated in situ
(they would rapidly loose all their energy traveling in the dense
photon environment close to the accretion disk),
therefore we require the jet to carry additional power to 
energize the emitting electrons (see in Fig. 1 the power corresponding 
to only the electron rest--mass, $L_{\rm e, cold}$). 
Note that also the distribution of the Poynting flux, $L_{\rm B}$,
has similar, but somewhat smaller, values than the radiative power.
This is a consequence of the large dominance of the high energy
peak shown by the SED of these sources, requiring the synchrotron 
being less important than the inverse Compton process, and then 
implying a relatively small value of the magnetic field.
For FSRQs we are then forced to consider other forms of jet power.
The simplest way to account for this ``missing power" is
to assume the presence of protons.
As can be seen, one proton per electron implies a jet power 
a factor 10--30 larger than what it radiates, i.e. a radiative
efficiency between 3 and 10 per cent.
This level of efficiency is what one expects in the ``internal shock"
scenario, and is also demanded by the requirement that most of the 
power should not be dissipated, since it must survive all along the 
jet and be deposited to the radio lobes.

There are alternatives.
The power could still be hidden in a component of the magnetic field
different from what exists in the emitting region.
This may appear contrived at first sight, but in reconnection regions
we may have indeed values of the magnetic field smaller than the average
(see e.g. Drenkhan \& Spruit 2002 for an application to GRBs).
Another possibility could be to have a very large amount of cold pairs.
But this is unlikely, since they would suffer i) strong annihilation at 
the base of the jet and ii) a strong Compton drag in the acceleration phase 
(see Celotti 2002).

\begin{figure}
\begin{tabular}{ll}
{\hskip -0.5 true cm
\psfig{figure=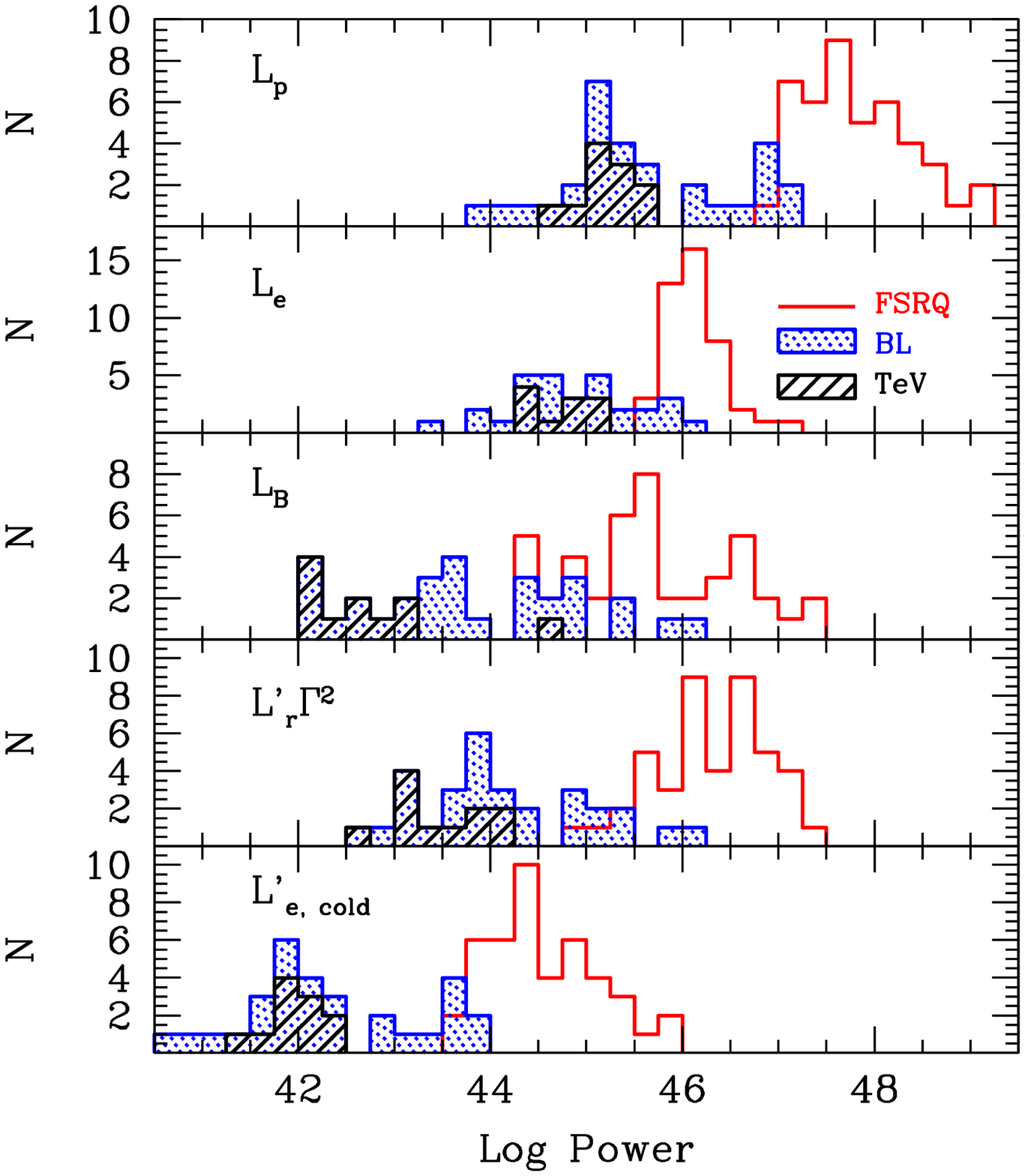,width=8.2cm}}&
{\hskip -1.5 true cm
\psfig{figure=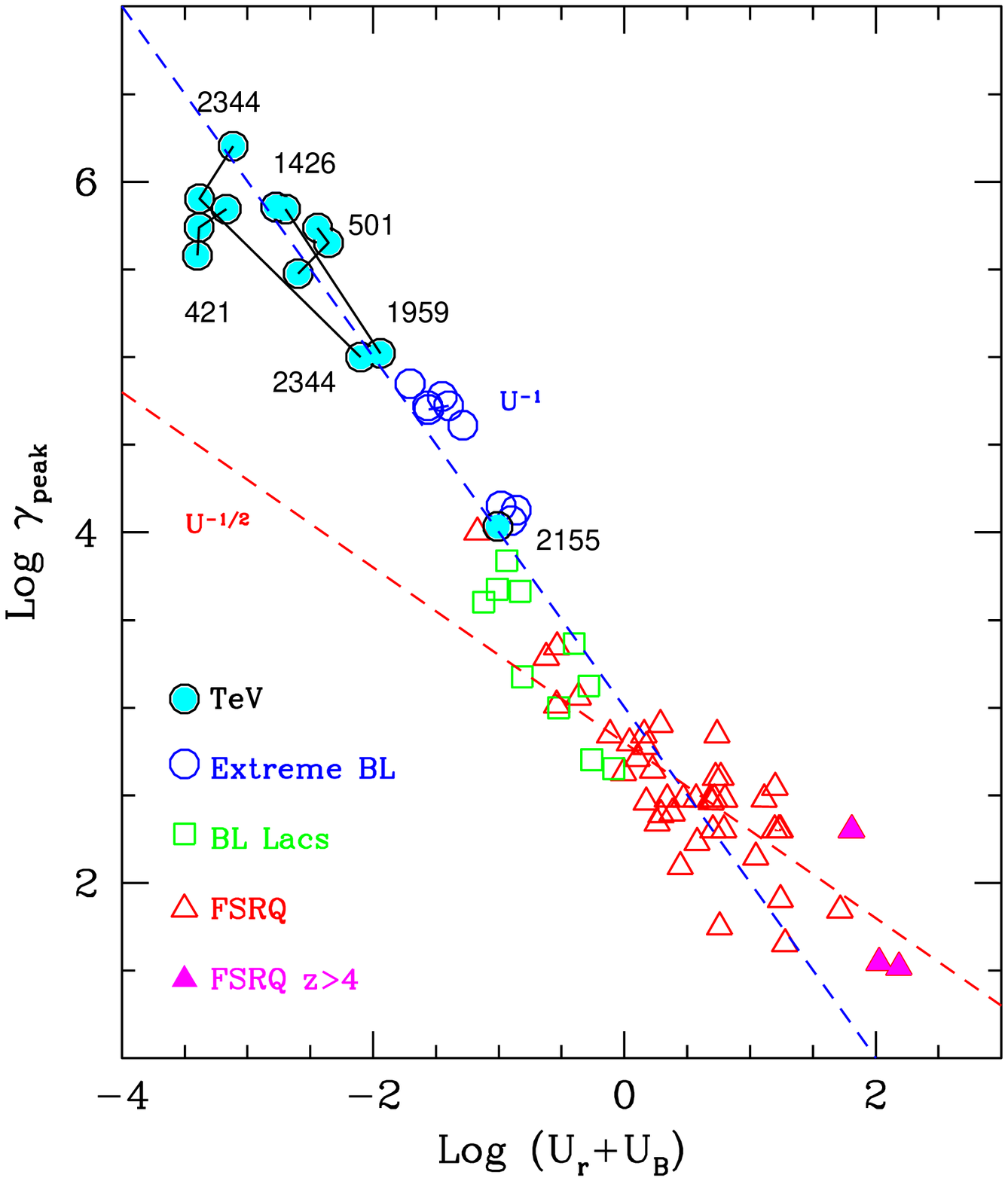,width=8.2cm}}\\
\end{tabular}
\caption{
{\bf Left:} Distribution of powers for $\gamma$--ray emitting 
blazars. $L_{\rm p}$ is the power carried by protons, assuming to have
one proton per emitting electron; $L_{\rm e}$ is the power carried by
the relativistic electrons, $L_{\rm B}$ is the Poynting flux;
$L^\prime_{\rm r}\Gamma^2\sim L_{\rm obs}/\Gamma^2$ 
is the power in the emitted radiation,
and $L_{\rm e, cold}$ is the power carried by the emitting electrons,
excluding their random energy. From Celotti \& Ghisellini, 2003, in prep.
{\bf Right:} The energy of the electrons emitting at the peaks of the SEDs,
$\gamma_{\rm peak}$, as a function of the comoving energy density 
(radiative plus magnetic) as seen in the comoving frame.
Note the presence of two branches (see text). From Ghisellini, 
Celotti \& Costamante (2002).
}
\label{fig1}
\end{figure}

\section{The blazar sequence: an update}

Fossati et al. (1998) suggested that the emitted luminosity 
controls the overall shape of the SED of blazars, while
Ghisellini et al. (1998), by applying a one--zone synchrotron
inverse Compton model to EGRET blazars could find a strong
correlation between the energy of the electrons emitting at the
peaks of the SED ($\gamma_{\rm peak}$) and the energy density $U$  
(radiative plus magnetic) in the comoving frame.
At those times the low power BL Lacs detected by EGRET were only
a few, with only Mkn 421 and Mkn 501 detected at TeV energies.
More recently, other 4 BL Lacs have been detected at 
TeV energies [namely 2155--304; 2344+514; 1426+428 and 1959+650,
Aharonian et al. 2001 and references therein; Holder et al. 2002].
These detections opened up the way to measure the poorly known far
IR cosmic background through its influence on the high energy
spectrum of TeV BL Blacs (see e.g. Stecker \& de Jager 1997). 
If the TeV emission we see is partially absorbed by the far IR,
it may be that all blazars (i.e. not only the powerful ones),
have their SED dominated by the $\gamma$--ray emission.
Also {\it Beppo}SAX made an important contribution to blazar
studies, particularly through the PDS instrument, sensitive between 
20 and 100 keV, detecting several BL Lacs in this band.
These recent observations allowed to enlarge the original sample of
EGRET blazars, and to explore the blazar sequence in the low power domain.
The right panel of Fig. 1 shows
that the correlation between $\gamma_{\rm peak}$ and $U$
is more complex than previously thought: it can be reproduced by assuming
that in low power BL Lacs we have $\gamma_{\rm peak} \propto U^{-1}$,
while for powerful blazars $\gamma_{\rm peak} \propto U^{-1/2}$,
which was the correlation found previously, considering only EGRET blazars.
This has been interpreted as the manifestation of a two--step acceleration
process, with electrons pre--accelerated at a certain value of $\gamma$ 
(which equals $\gamma_{\rm peak}$ for high power blazars), and then accelerated
at still higher energies, forming a broken power law distribution whose break 
is controlled by the radiative cooling after one light crossing time
(this break equals $\gamma_{\rm peak}$ in low power BL Lacs).


\section{Internal shocks and bulk Compton emission}

The result on the jet power and on the spectral modelling of blazars mentioned 
above can have a satisfactory explanation in the internal shock scenario, 
in which the central engine works intermittently producing blobs moving 
at slightly different velocities and therefore colliding at some distance 
from the black hole, transforming a few per cent of the bulk kinetic 
energy in radiation (see e.g. Ghisellini 1999; Spada et al., 2001).
The fact that most of the dissipation occurs at $\sim$hundreds of 
Schwartzchild radii is a natural consequence in this scenario 
(it is the distance of the first collisions between consecutive shells); 
variability is a built--in feature (but requires that the central engine 
works intermittently); acceleration of particles lasts for about one shell 
light crossing time, explaining the $\gamma_{\rm peak}-U$ correlation; 
dissipation of bulk kinetic energy into radiation occurs all along the jet, 
but with a decreasing efficiency, due to the decreasing contrast in the 
$\Gamma$--values of two colliding shells.
It can also account for the blazar sequence (Guetta et al. 2002)
with the additional very reasonable assumption that the broad line region
is located at a distance which correlates with the accretion disk luminosity
(which in turn correlates with the jet power):
in weak BL Lacs shell--shell collisions occur outside the broad line region,
in a much less dense external photon environment, implying less severe cooling
(i.e. large $\gamma_{\rm peak}$) and a relatively more important SSC emission.
It is also a relatively simple scenario, allowing quantitative analysis
(see Tanihata et al. 2002 for an application to Mkn 421).
Besides all that, part of the appeal of this scenario lies on the possibility
that $all$ relativistic jets work the same way, therefore including GBRs
and galactic superluminal sources.

On the other hand, as explained above and presented by Blandford (2002) and by 
Lyutikov (2002) at this meeting, the jet could be purely electromagnetic.
Electromagnetic instabilities could develop at some distance 
from the black hole, producing most of the blazar emission
without violating the $\gamma$--$\gamma$ constraints.
This could equally well happen in all relativistic jets.
We need thus a way to distinguish between the two scenarios,
bearing in mind that also in the internal shock scenario an acceleration
mechanism is required, which might use magnetic forces. 
However, the matter has to achieve its final bulk Lorentz factor quite 
rapidly, before the $\gamma$--ray dissipation zone.

One possibility to check for the presence of fastly moving large 
quantities of matter in the inner jet is through the interaction 
of this matter with the radiation produced by the accretion disk.
As Sikora et al. (1997) and Sikora \& Madejski (2002) pointed out,
even if the matter is cold in the comoving frame, bulk
Comptonization between jet matter and radiation can be a
very powerful tool to test the inner jet content.
The expected emission feature should be an X--ray excess at the frequency
$\nu_{\rm disk}\Gamma^2\sim$1 keV ($\nu_{\rm disk}$ is the peak frequency
of the disk radiation, in the UV band).
The level of this emission measures the optical depth of the jet, hence
its matter content.
Up to now there have been no claims of detection of this feature, which
however could be masked by the power law emission from the $\gamma$--ray
emission zone. Therefore there is the hope that more detailed observations
can either detect the feature or at least put strong upper limits on it.
Note that also purely electromagnetic models could accomodate for some
bulk Compton radiation, through some electron positron pairs created 
at the base of the jet, which would serve as scatterers without contributing
much to the total jet power. But in this case the level of the bulk Comptonization
feature is arbitrary. 
If instead the feature is detected at a level predicted by the bulk kinetic power 
estimated by other means, this should be considered as a severe problem
for the purely electromagnetic scenario.





\begin{thebibliography}{}


\bibitem[1]{a}Aharonian F. et al., 2001, 27th ICRC, astro--ph/0112314
\bibitem[2]{b}Blandford R.D., 2002, these proceedings
\bibitem[3]{c}Blandford R.D. \& Znajek R.L., 1977, MNRAS, 176, 465
\bibitem[4]{d}Celotti A. \& Fabian A.C. 1993, MNRAS, 264, 228
\bibitem[5]{e}Celotti A., 2002, these proceedings
\bibitem[6]{f}Drenkhahn G. \& Spruit H.C., 2002, A\&A, 391, 1141 
\bibitem[7]{g}Fossati G. et al., 1998, MNRAS, 299, 433 
\bibitem[8]{h}Ghisellini G. \& Madau P., 1996, MNRAS, 280, 67 
\bibitem[9]{i}Ghisellini G. et al., 1998, MNRAS, 301, 451 
\bibitem[10]{j}Ghisellini G., 1999, Astronomische Nachrichten, 320, p. 232 
\bibitem[11]{k}Ghisellini G. \& Celotti A., 2001, MNRAS, 327,  739 
\bibitem[12]{l}Ghisellini G., Celotti A. \& Costamante L., 2002, A\&A, 386,  833 
\bibitem[13]{m}Guetta D. et al., 2002, 5th AGN Italian Workshop, astro-ph/0210115 
\bibitem[14]{n}Holder J. et al., 2002, ApJL in press, astro--ph/0212170
\bibitem[15]{o}Lyutikov M., 2002, these proceedings, astro--ph/0211338
\bibitem[16]{p}Lyutikov M. \& Blandford R.D., 2002, astro--ph/0210671
\bibitem[17]{q}Rawlings S.G. \& Saunders R.D.E., 1991, Nature, 349, 138
\bibitem[18]{r}Sikora M., Madejski G., Moderski R. \& Poutanen J., 1997, ApJ, 484, 108
\bibitem[19]{s}Sikora M. \& Madejski G., 2002, astro--ph/0211587 
\bibitem[20]{t}Spada M. et al., 2001, MNRAS, 325, 1559 
\bibitem[21]{u}Stecker F.W. \& de Jager O.C., 1997, ApJ, 476,712
\bibitem[22]{v}Tanihata C. et al., 2002, ApJ in press, astro--ph/0210214 
\bibitem[23]{z}Tavecchio F. et al., 2000, ApJ, 544, L23

\end{thebibliography}
\end{document}